\documentclass[a4paper, 11pt]{article}
\usepackage{comment} 
\usepackage[nointegrals]{wasysym}
\usepackage{fullpage} 
\usepackage{graphicx}
\usepackage{cite}
\usepackage{multicol}
\usepackage[perpage]{footmisc} 
\usepackage{sectsty}
\sectionfont{\large}
\usepackage{graphicx}
\usepackage{subcaption}
\usepackage{bm}
\usepackage{float}
\usepackage{amsmath}
\usepackage[utf8]{inputenc}
\usepackage{amssymb}
\usepackage{mathrsfs}
\usepackage{wrapfig}
\usepackage{enumitem}
\usepackage[english]{babel}
\usepackage{comment}
\usepackage{bbold}  
\usepackage[usenames,dvipsnames]{xcolor} 
\usepackage{xcolor,hyperref}
\usepackage{setspace}
\usepackage{abstract}
\usepackage[font=footnotesize,labelfont=bf]{caption}
\usepackage{tablefootnote}

\usepackage[T1]{fontenc}
\usepackage[margin=0.94in]{geometry}
\def\Mpl{M_{\rm P}}

\usepackage{enumitem}
\usepackage{graphicx}
\usepackage{subcaption}
\usepackage{xcolor}
\usepackage{amssymb}

\numberwithin{equation}{section}

\begin{document}
\null\hfill {MPP-2026-1\\ \null\hfill IPMU26-0001\\ \null\hfill YITP-26-04} \\
\vspace*{\fill}

\vspace*{\fill}
\begin{center}
    \LARGE\textbf{{ \textcolor{Black}{The \textit{recipe} for the degrees of freedom.
    }} }

    \normalsize\textsc{Anamaria Hell$^{1}$, Elisa G. M. Ferreira$^{1}$, Dieter Lüst,$^{2,\;3}$, Misao Sasaki$^{1,4,5,6}$}
\end{center}

\begin{center}
     $^{1}$ \textit{Kavli IPMU (WPI), UTIAS,\\ The University of Tokyo,\\ Kashiwa, Chiba 277-8583, Japan}
    \\
     $^{2}$\textit{Arnold Sommerfeld Center for Theoretical Physics,\\
Ludwig–Maximilians–Universität München\\
Theresienstraße 37, 80333 Munich, Germany}\\
$^{3}$\textit{Max–Planck–Institut für Physik (Werner–Heisenberg–Institut)\\
Boltzmannstra{\ss}e 8, 85748 Garching, Germany}\\ 
    $^{4}$ \textit{Center for Gravitational Physics and Quantum Information,\\ Yukawa Institute for Theoretical Physics, Kyoto University, Kyoto 606-8502, Japan}\\
     $^{5}$ \textit{Leung Center for Cosmology and Particle Astrophysics,\\ National Taiwan University, Taipei 10617, Taiwan}\\
  $^{6}$ \textit{ Asia Pacific Center for Theoretical Physics,\\ Pohang 37673, Korea}
\end{center}
\thispagestyle{empty}

\renewcommand{\abstractname}{{\textcolor{Black}{Abstract}}}
\begin{abstract}
    We consider the question of counting the degrees of freedom in theoretical models, with an emphasis on theories of fields and gravity. Among the possible approaches, the Hamiltonian formulation remains one of the most systematic and robust tools. However, it can easily become long and technically involved.  In this work, we present a broadly applicable recipe to find the degrees of freedom directly, based on the Lagrangian formulation. We compare it to the standard approaches, highlight the challenges that may arise in the latter, and demonstrate that the proposed method leads to transparent insights about the dynamical nature of theory in a quick, simple, and straight-forward way. 
\end{abstract}

\vfill
\small
\noindent\href{mailto:anamaria.hell@ipmu.jp}{\text{anamaria.hell@ipmu.jp}}\\
\href{mailto:elisa.ferreira@ipmu.jp}{\text{elisa.ferreira@ipmu.jp}}\\
\href{luest@mppmu.mpg.de}{luest@mppmu.mpg.de}\\
\href{mailto:misao.sasaki@ipmu.jp}{\text{misao.sasaki@ipmu.jp}}
\vspace*{\fill}

\clearpage
\pagenumbering{arabic} 
\newpage

\tableofcontents
\newpage
\section{Introduction}

One of the most fundamental questions that can be asked when studying any physical theory is: \textit{What are its degrees of freedom?} Mathematically, they are determined by the number of initial conditions necessary to specify a well-posed Cauchy problem of the equations of motion of the system, divided by two. This concept can be found on the first pages of standard literature in the course of \textit{"Mechanics"}, and is especially relevant\cite{llandau60:mech}. The degrees of freedom (dof) serve as building blocks of the models that aim to describe processes in our Universe, telling us key information about their dynamics, validity, and connections to observations. 

Among the approaches to find the propagating modes, the Dirac-Bergmann algorithm has been particularly useful \cite{Bergmann:1949zz, Dirac:1950pj, Anderson:1951ta, Dirac:1958sc, DiracC}. This method, rooted in the Hamiltonian formalism, classifies all of the constraints in a theory by using the Poisson or Dirac brackets and subtracts them from the total number of possible modes to get the number of physical degrees of freedom. Another approach, also related to the Hamiltonian formalism, is the Faddeev-Jackiw method \cite{Faddeev:1988qp}, which reformulates the system by using the first-order action -- a form containing only the first-order time-derivatives of the fields from which the Hamiltonian density is subtracted -- solving the constraints explicitly, and substituting them back into the action to obtain a reduced expression. Both of these methods rely on the Hamiltonian formulation. However, one may also choose to instead analyze the field equations directly. This formally rigorous approach originates in the works of É. Cartan, A. Einstein, M. Kuranishi, and W. M. Seiler, \cite{Cartan1, Cartan2, Einstein1, Kuranishi, EC, Seiler1, Seiler2}, and has been recently also extended to modified theories of gravity \cite{Heisenberg:2025fxc}, with a particular emphasis on the $f(Q)$ gravity. In essence, it relies on an idea of using the formal power series expansion, applied directly to the partial differential field equations. 

All of these approaches are reliable and applicable for a wide range of theoretical models. Moreover, the Hamiltonian formalism is particularly powerful in gravitational theories, where the space-time is curved, as it provides a background-independent determination of the number of degrees of freedom via the Arnowitt-Deser-Misner (ADM) procedure \cite{Arnowitt:1962hi}. At the same time, applying these methods often leads to lengthy and technically involved calculations. This issue becomes especially important in theories beyond Einstein's General Relativity, or the Standard Model of Particle 
Physics. While simple extensions, such as adding the fields a mass term, or another interaction can typically be easily handled, the situation drastically changes when one involves gravity, or non-linear extensions of gauge fields: the calculations become rapidly non-trivial, and the full counting of the propagating modes requires a substantial work, while the nature of the degrees of freedom may appear often somewhat obscure. 

To avoid these issues, a common strategy is to rely on the Lagrangian formulation and analyze a theory in a manifestly covariant manner. This often involves choosing the manifestly covariant gauges --  such as a Lorentz gauge, in the presence of gauge-redundancy -- or applying a Stueckelberg trick, which introduces additional degrees of freedom such that the resulting theory acquires a gauge redundancy \cite{Ruegg:2003ps, Stueckelberg:1938hvi}. Originally developed for the  Proca theory -- a theory of a massive vector field -- the Stueckelberg trick has also been generalized to other gauge theories, including massive gravity and non-Abelian fields \cite{Dvali:2006su, deRham:2014zqa, Hinterbichler:2011tt, Kunimasa, Vainshtein, Gambuti:2020onb, Huang:2007xf, Dvali:2007kt, deRham:2011qq}.  In gravitational theories, it involves adding scalar and vector degrees of freedom, followed by imposing the Lorentz-like gauge conditions, with vanishing 4-divergence of the tensor and vector modes (see eg \cite{Hinterbichler:2015soa, Alvarez:2018lrg, Dalianis:2020nuf, Kamimura:2021wzf,  Kubo:2022jwu, Kubo:2022dlx, Kubo:2022lja, Buoninfante:2023ryt, Ghosh:2023gvc, Karananas:2024hoh}).  As a result, the relations between the old and the new variables resemble helicity decomposition. 

However, such an approach may lead to fictitious modes -- modes that are an artifact of gauge redundancy. As demonstrated on an example of pure $R^2$ gravity in flat spacetime in \cite{Hell:2023mph}, the Lorentz-like gauges do not fix the gauge entirely. Thus, if one studies a theory in a manifestly covariant way, as in \cite{Alvarez-Gaume:2015rwa}, and compares it to studying the fields directly \cite{Hell:2023mph}, one could find contradictory results on the number of degrees of freedom -- one yields a (fictitious) scalar mode which is absent in the second approach. The latter strategy is rooted in studying the constraints and the propagating fields directly, without any introduction of gauge redundancy. This method is based on the idea of cosmological perturbation theory \cite{Mukhanov:1981xt, Chibisov:1982nx,   Kodama:1984ziu, Sasaki:1986hm, Mukhanov:1990me}, where the metric is decomposed according to the irreducible representations of the rotation group, into scalar, vector, and tensor modes. 

Studying theories in a way that is not manifestly Lorentz covariant is extensively represented in the standard literature of quantum field theory, although not excluded  \cite{Schwinger:1948iu, Schwinger:1948yk, Boulware:1962zz, Deser:1972wi}. Nevertheless, it is particularly useful -- it allows one to have a clear look into the building blocks of theory, expressing the action only in terms of the propagating modes.  This approach has been crucial in investigating the massless limit of massive gauge theories, and the effects of the non-linear terms \cite{Chamseddine:2012gh, Chamseddine:2018gqh, Hell:2021wzm, Hell:2021oea, Hell:2022wci, Hell:2024xbv, Hell:2025uoc, Hell:2025pso}. It is also almost essential in studying the stability of various modified theories of gravity -- while it does not exclude the Hamiltonian formalism, it allows one to a more clear view on how different degrees of freedom scale with time, behave in various limits, and are potentially problematic (see eg. \cite{Demozzi:2009fu, Alberte:2010it, Alberte:2010qb,Emami:2016ldl, Tsujikawa:2022aar,  DeFelice:2023psw, Capanelli:2024pzd, DeFelice:2024seu, Bahamonde:2024zkb, DeFelice:2025ykh, Hell:2025lgn, Hell:2025wha, DeFelice:2025khe, Chen:2025aom, Hell:2025lbl}). It is also especially useful for analysis of perturbations around concrete backgrounds that give rise to a different number of propagating modes, and are significantly more obscure in the Hamiltonian formalism \cite{Hell:2023mph, Hell:2025lbl, Barker:2025gon}. 

Despite the usefulness of this approach, to date, such studies have so far taken place only on a case-by-case basis, and a general prescription underlying this method remains unknown. In this work, we will fill this gap. Starting from the basic examples such as a free particle in classical mechanics, we will develop a systematic recipe for finding the degrees of freedom, expressing the action only in terms of the propagating modes, which will be applicable across a broad class of theoretical models.

This work is structured as follows: first, in Section 2, we will cover the basics of the degrees of freedom and provide basic definitions on simple examples. 
In Section 3, we will discuss the main challenges that one encounters when studying theories through standard, more conventional approaches. In Section 4, we will formulate the recipe,  demonstrate it in Section 5 on a concrete example, and compare it with the Hamiltonian approach in  Section 6. Finally, in Section 7, we will summarize our findings.

\section{The basics of the DOF}

In this section, we will lay the foundations to study the degrees of freedom (dof), staring from the simplest concepts on mechanics, and building up to the classical field theory. To keep this work accessible to a broad audience, we will focus on elementary examples. Therefore, readers who are already familiar with the basic notions of field theory may wish to skip to Section 3, where we discuss the main challenges and more involved examples.

\subsection{Particles and fields}

The simplest possible case of a physical system is a \textit{free particle}. In (1+1) dimension\footnote{This corresponds to one time and one space dimension.}, this object is described by one coordinate, $q$, satisfying the following equation of motion:
\begin{equation}
    \Ddot{q}(t)=0, 
\end{equation}
where the dot denotes the derivative with respect to time, and whose general solution is given by:
\begin{equation}
    q(t)=At+B,
\end{equation}
where $A$ and $B$ are constants of integration. The above solution describes trajectories that a free particle can make. However, if we wish to connect it to the motion of an object in the real world, we should specify its initial position and initial velocity:
\begin{equation}
    q(t_0)\qquad\text{and}\qquad \dot{q}(t_0),
\end{equation}
where $t_0$ is the initial moment of time. The \ textit { number of degrees of freedom (dof)} $s$ is then defined as the number of initial conditions necessary to fully specify the Cauchy problem of the differential equation that the physical object satisfies, divided by two. In the case of the above free particle in one dimension, this number is:
\begin{equation}
    \frac{2  \text{ initial conditions}}{2}=1 \textit{ dof }.
\end{equation}
For a single particle, a generalisation to three spatial dimensions (or more) is then straightforward. 
In this case, its position is characterised by a three (or d-) dimensional vector $\Vec{q}(t)$, with each of its components satisfying:
\begin{equation}
    \Ddot{\vec{q}}\;(t)=0
\end{equation}
Clearly, we now have to specify three (or d-) pairs of initial conditions $\{\vec{q}(t_0),\;\dot\vec{q}(t_0)\}$. Thus, in this case, the number of degrees of freedom in d-spatial dimensions will be $d$. Moreover, instead of a single particle, we can also focus on several of them.  $N$ of them, in d-dimensions, obey the following equation of motion:
\begin{equation}
    \Ddot{q}_i(t)=0,\qquad \text{where},\qquad i=1\; ... \;Nd,
\end{equation}
and corresponds to $2 Nd$ initial conditions. The above procedure can be naturally generalised to scalar fields \cite{Mukhanov:2007zz}. Instead of labelling each particle and its corresponding position with a discrete index $i$, we can promote it to a continuous label, $\vec{x}$, which would correspond to the position in space. The scalar field can then be thought of as a simple generalisation of a particle:
\begin{equation}
    q_i(t)\qquad \to\qquad q_{\vec{x}}(t),\qquad \text{by setting}\qquad i\to\vec{x},
\end{equation}
taking its values at all points in space-time. Therefore, in simplest terms, one could write the following equation for a scalar field:
\begin{equation}
    \Ddot{q}_{\vec{x}}(t)\equiv\Ddot{q}(t,\vec{x})=0,
\end{equation}
and conclude that it describes an infinite number of dof. One should note that the above relation violates Lorentz invariance. The scalar field of a Lorentz invariant theory instead satisfies:
\begin{equation}\label{healthyscalareom}
    \Box\phi(t,x)=0,
\end{equation}
where $\Box=-\partial_0^2+\partial_i\partial_i$ is the D'Alambertian (wave) operator. The above equation also describes an infinite number of dof -- the initial value problem requires us to specify all points in space for a given time $t_0$. However, in the case of more complicated fields, it is more natural to speak about the degrees of freedom per point in space-time.  Thus, we say that the above scalar field describes a single dof (and omit the wording -- per point in space-time). 

One can speak about the notion of the dof by only focusing on the equations of motion. When we construct new theories or analyse more complicated ones, we usually work in either Lagrangian or Hamiltonian formalism. Both of these approaches are equivalent and are used to derive the equations of motion of the system. In the following, we will provide the very basics of the two approaches in a nutshell, followed by a more general and standardised field theory approach in Section 3. 
\subsection{Mechanics in a nutshell}

In the previous subsection, we have formulated the notion of dof for a free particle and generalized it to the scalar field. This notion is closely related to the equation of motion, because one determines its number depending on the initial conditions. In practice, one most commonly formulates a physical system using the Lagrangian or Hamiltonian approach, from which the equations of motion follow. In this section, we will focus on these two approaches in the simplest cases -- a single free particle and the scalar field. 

\subsubsection{The variational principle}
The action for a particle, characterized with the generalized coordinate $q$ and its generalized velocity $\dot{q}$, is defined as a functional:
\begin{equation}
    S[q]=\int dt L(q(t), \dot{q}(t)),
\end{equation}
where $L$ is the Lagrangian of the particle. In the free case, it is simply given by:
\begin{equation}\label{freePL}
    L=\frac{1}{2}\dot{q}^2. 
\end{equation}
To find the equations of motion, one simply varies the above action with respect to the generalized coordinate. Then by requiring that the action is minimal, 
\begin{equation}
\delta S=0,
\end{equation}
one finds the equations of motion. If we want to consider many more particles, we simply sum the action related to them. For example, for free N particles: 
\begin{equation}
    S[q_1,\; ...,\; q_N]=\sum_{i=1}^N\int dt L(q_i(t), \dot{q}_i(t)),
\end{equation}
where $L$ is given by (\ref{freePL}), although we can also consider the possibility that the particles interact with each other as well. 

Following the line of the previous subsection, it is now easy to generalize this to a scalar field by replacing $i\to \vec{x}$. Then, the action for a free Lorentz invariant scalar field is given by:
\begin{equation}
    S[\phi]=\int d^3x\int dt \mathcal{L}(\phi(t,x), \dot{\phi}(t,x)). 
\end{equation}
where $\mathcal{L}$ is the corresponding Lagrangian density. Similarly, by varying the above action with respect to the scalar field, one finds the corresponding equations of motion. 
As an example, we can consider a free scalar field:
\begin{equation}\label{LagrDenScalarFree}
    \mathcal{L}=-\frac{1}{2}\partial_{\mu}\phi\partial^{\mu}\phi,
\end{equation}
where $\partial_{\mu}$ is the partial derivative with respect to the $x^{\mu}$ coordinate. In particular\footnote{We will be using the (-,+,+,+) signature. }:
\begin{equation}
    \mathcal{L}=\frac{1}{2}\dot{\phi}\dot{\phi}-\frac{1}{2}\phi_{,i}\phi_{,i},
\end{equation}
where $\phi_{,i}\equiv\partial_i\phi$,  $\partial_i=\partial^i$, and $\partial_0=-\partial^0$. 
\subsubsection{The Hamiltonian approach}
An equivalent approach to the Lagrangian formalism is the Hamiltonian approach. While the physical  system in the former is determined by the generalized coordinates/fields and their corresponding generalized velocities, the Hamiltonian formalism replaces the velocities with the canonical momenta, defined by:
\begin{equation}
  p_i=\frac{\partial L}{\partial \dot{q}_i}, \qquad \text{for}\qquad i=1,\;...\;N
\end{equation}
for N particles, or, for scalar fields:
\begin{equation}
    \pi=\frac{\partial\mathcal{L}}{\partial\dot{\phi}}. 
\end{equation}
Then, by finding the relation between the momenta and the generalized velocity, and inverting this relation:
\begin{equation}
    \dot{\phi}=\dot{\phi}(\pi,\phi),
\end{equation}
One finds the Hamiltonian for N particles:
\begin{equation}
    H=\sum_{i=1}^N\left(p_i\dot{q}_i(p_i,q_i)-L(q_i, \dot{q}_i(p_i,q_i))\right),
\end{equation}
and, in more details, the Hamiltonian density $\mathcal{H}$ for the scalar field:
\begin{equation}
    H=\int d^3x\mathcal{H},\qquad \mathcal{H}=\pi\dot{\phi}(\pi,\phi)-\mathcal{L}.
\end{equation}
For a scalar field, we then find:
\begin{equation}
    H=\frac{1}{2}\pi^2+\frac{1}{2}\phi_{,i}\phi_{,i}
\end{equation}
In contrast to the Lagrangian formalism, the equations of motion in the Hamiltonian formulation are of first order. For a scalar field, they are given by:
\begin{equation}
    \dot{\pi}=\Delta\phi\qquad \text{and}\qquad \dot{\phi}=\pi
\end{equation}
To specify the Cauchy problem for the above equations, we again need two initial conditions. Therefore, we find a single dof. 

In this section, we have introduced the foundations of any field theory or theory of gravity. In the following, we will build on these concepts further. For a reader who wishes to go more in depth, and also discuss quantization, we reccomend studying the first chapters of \cite{Mukhanov:2007zz}, or other standard literature.

\section{The main challenges}

In the previous section, we have introduced the basic notions in theoretical physics using the simplest possible theories -- a free particle and a scalar field. However, many physical phenomena in nature, such as the propagation of light, require more intricate theoretical frameworks. In this section, we will discuss some of the main challenges that occur when studying these systems, and set the stage for overcoming them in Section 4. 

\subsection{Constrained systems}

The first challenge that one might imagine when counting the number of dof is to replace the scalar fields with the vector ones. These fields have four components in four dimensions: $A_{\mu}$, where $\mu$ runs from 1 to 3. So, naively, one might suspect that the following action in flat space-time describes four dof:
\begin{equation}\label{actProca}
    S=\int d^4x\left[-\frac{1}{4}F_{\mu\nu}F^{\mu\nu}-\frac{m^2}{2}A_{\mu}A^{\mu}\right],
\end{equation}
where 
\begin{equation}\label{fstrT}
    F_{\mu\nu}=A_{\nu,\mu}-A_{\mu,\nu},
\end{equation}
is the field-strength tensor. Nevertheless, to be fully sure, we should study the time derivatives, and on which vector field components do they act more closely. The above theory is known as the Proca theory, and it describes the dynamics of a massive vector field \cite{Proca:1936fbw}.  To find its dof, will follow the procedure similar to \cite{Demozzi:2009fu}. First, we will decompose the vector field into temporal and spatial components:
\begin{equation}\label{vectempspd}
    (\;A_0,\;A_i\;). 
\end{equation}
Due to the spherical symmetry, we can further decompose the spatial part of the vector field as:
\begin{equation}\label{vecdeclin}
    A_i=A_i^T+\chi_{,i},\qquad \text{where}\qquad A_{i,i}^T=0. 
\end{equation}
This is a decomposition into irreducible parts of the rotation group, also known as the Helmholtz decomposition. In the following, we will refer to the vectors $A_i^T$ as transverse modes, and $\chi$ as the longitudinal mode. With this decomposition, the action (\ref{actProca}) becomes:
\begin{equation}
    \begin{split}
       S=\frac{1}{2}\int d^4x&\left[A_0(-\Delta+m^2)A_0-2A_0\Delta\dot{\chi}-\dot{\chi}\Delta\dot{\chi}-m^2\chi\Delta\chi+\dot{A}_i^T\dot{A}_i^T-A_{i,j}^TA_{i,j}^T-m^2A_i^TA_i^T\right]
    \end{split}
\end{equation}
We can notice that among the above fields, the $A_0$ component is special -- it does not appear with any time derivatives. This is known as a \textit{constrained field}\footnote{In the Hamiltonian formalism, it leads to the primary constraint, meaning that its conjugate momentum vanishes, which then further leads to secondary constraints.}. Therefore, not all components of the vector field are propagating. To find the actual number of propagating modes, let us first find the constraint. By varying the action with respect to $A_0$, we find: 
\begin{equation}
    (-\Delta+m^2)A_0=-\Delta\dot{\chi}
\end{equation}
By solving the above equation for $A_0$, and substituting it back into the action, we arrive at the following Lagrangian density:
\begin{equation}
    \mathcal{L}=-\frac{1}{2}\chi(-\Box+m^2)\frac{-\Delta m^2}{-\Delta+m^2}\chi-\frac{1}{2}A_i^T(-\Box+m^2)A_i^T.
\end{equation}
The above Lagrangian density describes three degrees of freedom. The longitudinal mode satisfies the following equation of motion:
\begin{equation}
    (-\Box+m^2)\chi=0.
\end{equation}
Clearly, we need two initial conditions (per point in spacetime) to determine its propagation, which leads to one degree of freedom. In addition to the longitudinal mode, we have the transverse modes, which satisfy:  
\begin{equation}
    (-\Box+m^2)A_i^T=0. 
\end{equation}
This equation of motion gives rise to two additional independent dof, and not three, since one of the components of the transverse modes is related to the other two due to $A_{i,i}^T=0$. 

Therefore, we can see that as soon as one departs from the standard canonical theory of a  scalar field, one can reach a challenge -- if one describes theories of vector and tensor modes, or even p-forms, one has to additionally deal with constraints -- equations for components which do not propagate, but are functions of other fields, and are possible to identify by the absence of a kinetic term. 

\subsection{Gauge redundancy}

A next challenge that one might encounter in various theories is the presence of gauge redundancy -- invariance of the action under local transformations of fields. The most famous example of this type of theory is \textit{electrodynamics}, whose action is given by:
\begin{equation}
    \mathcal{L}=-\frac{1}{4}F_{\mu\nu}F^{\mu\nu},
\end{equation}
where the field strength tensor is defined in (\ref{fstrT}). We can notice that if one performs the field transformation:
\begin{equation}\label{U1gt}
    A_{\mu}\to\Tilde{A}_{\mu}=A_{\mu}+\partial_{\mu}\lambda,
\end{equation}
then
\begin{equation}
    F_{\mu}\to\Tilde{F}_{\mu\nu}=F_{\mu\nu}
\end{equation}
and thus, the action of electrodynamics is invariant under the (local) U(1) gauge transformation (\ref{U1gt}). 

One of the standard approaches to count the dof in this theory relies on based on using this gauge redundancy while keeping the analysis manifestly Lorentz-covariant. In particular, we can notice that  (\ref{U1gt}) allows us to pick $\lambda$ such that 
\begin{equation}
    \partial^{\mu}\Tilde{A}_{\mu}=0. 
\end{equation}
This makes one of the four components of the vector field $A_{\mu}$ dependent on the other three. This gauge choice is possible if the gauge parameter satisfies:
\begin{equation}
    \Box\lambda=-\partial^{\mu}A_{\mu}. 
\end{equation}
However, the above equation has a special case, when $A_{\mu}=0$. This leads to additional gauge redundancy in the theory:
\begin{equation}
    \Tilde{A}_{\mu}\to\Bar{A}_{\mu}= \Tilde{A}_{\mu}+\partial_{\mu}\gamma,
\end{equation}
where the gauge parameter $\gamma$ satisfies:
\begin{equation}
    \Box\gamma=0. 
\end{equation}
We can further use this to make one additional component of the vector field dependent on the other ones, leaving us thus with only two independent components of the vector field $\Bar{A}_{\mu}$ -- the two degrees of freedom, since the equations of motion are second order:
\begin{equation}
    \partial^{\mu}F_{\mu\nu}=0. 
\end{equation}

An alternative way to count the degrees of freedom is following the approach that we presented in the previous subsection. By decomposing the vector field into its temporal and spatial components, and further using (\ref{vecdeclin}), we find the following Lagrangian density:
\begin{equation}\label{edLcomp}
       \mathcal{L}=\frac{1}{2}\left[-A_0\Delta A_0-2A_0\Delta\dot{\chi}-\dot{\chi}\Delta\dot{\chi}+\dot{A}_i^T\dot{A}_i^T-A_{i,j}^TA_{i,j}^T\right].
\end{equation}
We can notice that the above case is very similar to the massive theory -- the temporal component $A_0$ is still not propagating. By varying the action with respect to it, we find:
\begin{equation}
    \Delta A_0=\Delta\dot{\chi},
\end{equation}
which leads us to
\begin{equation}
    A_0=\dot{\chi}. 
\end{equation}
By substituting this back into (\ref{edLcomp}), we find that the scalars cancel, leaving us only with two independent vector modes: 
\begin{equation}\label{edLcomp}
       \mathcal{L}=\frac{1}{2}\left[\dot{A}_i^T\dot{A}_i^T-A_{i,j}^TA_{i,j}^T\right],
\end{equation}
which correspond to the two polarisations of the photon. 

It is useful to note at this point to notice that while $\Bar{A}_{\mu}$ was depending on the choice of the gauge\footnote{Eg. one could have chosen another $\lambda$, for example, choosing it in such a way that corresponds to the Coulomb gauge, or temporal gauge.}, there was only one way to reach the $A_i^T$ component, by resolving the constraints. Notably, $A_i^T$ component is \textit{gauge-invariant} -- it is unchained by the transformations (\ref{U1gt}). It is related to the original vector field $A_i$ through the spatial projector:
\begin{equation}
    A_i^T=P_{ij}^TA_j=\left(\delta_{ij}-\frac{\partial_i\partial_j}{\Delta}\right)A_j
\end{equation}
which vanishes when it encounters a particular spatial derivative:
\begin{equation}
    P_{ij}^T\partial_j\lambda=0. 
\end{equation}

\subsection{When the standard approach fails}

Electrodynamics is one of the simplest gauge theories. Beyond it, there are various other theories, with more emerging every day, driven by the numerous open cosmological puzzles and an ongoing quest to find a theory that could resolve them. These theories can be far from simple. Therefore,  one must carefully choose which formalism or approach is best suited for their analysis. Depending on the choice, the length of the computation can greatly increase, and one can encounter significantly more complicated expressions, from which it is difficult to interpret what the theory actually describes. 

With this in mind, one could be tempted to study theories in a manifestly Lorentz-covariant way, as we have previously described. The Hamiltonian formalism usually involves a background independent study especially in the gravitational systems, whereas resolving the constraints can easily lead to huge number of terms\footnote{See for example \cite{DeFelice:2025ykh}, where just the starting point for the vector fields with non-minimal coupling to gravity involved 25 terms in the Lagrangian just for scalar perturbations in the FLRW Cosmology, and 95 terms for the even modes for the Bianchi Type 1 Universe.} If there is an easier way, then it would be natural to follow it. The standard approach described in the previous section seems to fit into this category, as it keeps the number of terms minimal. However, it appears that it is not always reliable. 

To demonstrate this, let us consider another theory of a vector field, described by the following Lagrangian:
\begin{equation}\label{3formvec}
    \mathcal{L}=\frac{1}{2}\left(\partial_{\mu}A^{\mu}\right)^2. 
\end{equation}
Let us analyze this theory along the lines of the previous approach. We first write the vector field as \footnote{Writing the field in this way is usually reoffered to decomposing the field  \textit{a la Stueckelberg.}  However, one should not regard it as an decomposition, but an introduction of an extra scalar mode, as pointed out in \cite{Hinterbichler:2011tt}.  }:
\begin{equation}
    A_{\mu}=W_{\mu}+\partial_{\mu}\phi. 
\end{equation}
This automatically introduces a gauge redundancy:
\begin{equation}
    W_{\mu}\to\Tilde{W}_{\mu}+\partial_{\mu}\lambda\qquad\text{and}\qquad  \phi\to\Tilde{\phi}=\phi-\lambda
\end{equation}
We can choose the parameter $\lambda$ to set
\begin{equation}\label{Lguage}
    \partial_{\mu}\Tilde{W}^{\mu}=0,
\end{equation}
after which, the Lagrangian becomes:
\begin{equation}
    \mathcal{L}=\frac{1}{2}\Box\phi\Box\phi. 
\end{equation}
This implies that there are two degrees of freedom, since there are four derivatives acting on $\phi$. However, the previous gauge choice also leads us to the residual gauge redundancy:
\begin{equation}\label{resG}
     W_{\mu}\to\Tilde{W}_{\mu}+\partial_{\mu}\gamma\qquad\text{and}\qquad  \phi\to\Tilde{\phi}=\phi-\gamma
\end{equation}
where $\gamma$ satisfies:
\begin{equation}
    \Box\gamma=0. 
\end{equation}
We can choose it to fix an additional mode. Therefore, the above analysis leads us to conclude that the free theory described by Lagrangian (\ref{3formvec}) describes one degree of freedom. 

If instead we choose to study the constraints directly in the Lagrangian, we find the following expression, upon decomposing the vector field according to (\ref{vectempspd}) and (\ref{vecdeclin}):
\begin{equation}\label{3formvec2}
    \mathcal{L}=\frac{1}{2}\left(\dot{A}_0-\Delta\chi\right)^2. 
\end{equation}
Therefore, in contrast to Proca or Maxwell theory, the $A_0$ component is now propagating, while the longitudinal mode $\chi$ is constrained with the following relation:
\begin{equation}\label{chiconstr}
      \Delta\chi=\dot{A}_0. 
\end{equation}
By expressing the mode $\chi$ in terms of the temporal component according to the above constraint, we find that the Lagrangian density (\ref{3formvec2}) becomes: 
\begin{equation}
    \mathcal{L}=0, 
\end{equation}
implying thus that there are no degrees of freedom, as pointed out in \cite{Hell:2023mph}. At this stage, one might object that if one instead takes the scalar field 
\begin{equation}
    \mathcal{L}=\frac{1}{2}\phi\Box\phi, 
\end{equation}
and substitutes the equation of motion 
\begin{equation}\label{scalarfieldeom}
    \Box\phi=0 
\end{equation}
into the Lagrangian density, one would find that it vanishes as well. But, this clearly wouldn't imply that the number of dof for the scalar field is zero. Therefore, how could one conclude it from the previous example of the vector field? The crucial difference between (\ref{scalarfieldeom}) and (\ref{chiconstr}) is that the former is the equation of motion, while the latter is a constraint -- an equation that relates the two fields, analogous to the constraints in the Hamiltonian formalism. 

The main reason for the discrepancy between the direct and the manifestly covariant approach lies in the gauge redundancy, which is not fully fixed in the latter one.  In other words, the scalar $\phi$ that seemed to propagate in the manifestly covariant approach was a \textit{fictitious mode} --  an artifact of the bad gauge choice. 
If one chooses the Lorentz gauge, as in (\ref{Lguage}), we have 
\begin{equation}
    \Box\lambda=\partial^\mu W_\mu\,.
\end{equation}
This implies that $\lambda$ contains a scalar field degree of freedom, as the above can be satisfied by  $\bar\lambda=\lambda+\gamma$ for any choice of $\gamma$ satisfying $\Box\gamma=0$.

The appearance of this degree of freedom simply means that the Lorentz gauge (\ref{Lguage}) does not completely fix the gauge, as we have seen in (\ref{resG}).
In contrast, if one chooses the Coulomb gauge, for which
\begin{equation}
    \partial_{i}\Tilde{W}_i=0,
\end{equation}
we have $\partial^i\partial_i\lambda=\partial^iW_i$, which fixes $\lambda$ uniquely. 

It is worth noting that the theory of the vector field that we have previously discussed is the theory of a 3-form in disguise:
\begin{equation}
    \mathcal{L}=-\frac{1}{48}H_{\mu\nu\rho\sigma}H^{\mu\nu\rho\sigma}, 
\end{equation}
where 
\begin{equation}
    H_{\mu\nu\rho\sigma}=C_{\nu\alpha\beta,\mu}-C_{\mu\alpha\beta,\nu}+C_{\beta\mu\nu,\alpha}-C_{\alpha\mu\nu,\beta}, 
\end{equation}
is the field-strength tensor of a 3-form, $C_{\mu\nu\rho}$, which is totally antisymmetric, and related to the previous vector field with the Levi-Civita tensor: 
\begin{equation}
       A_{\mu}=\varepsilon_{\mu\nu\rho\sigma}C^{\nu\rho\sigma}.
\end{equation}
It is well known that this theory propagates no modes if it is massless. Therefore, this further confirms that the direct approach yielded a correct result. 

A similar case to the previous discussion was also found for the pure $R^2$ gravity, described by the action:
\begin{equation}
    S=\int d^4x\sqrt{-g}R^2, 
\end{equation}
in flat space-time. If one uses the manifestly covariant approach with:
\begin{equation}\label{decompositionMCOV}
h_{\mu\nu}=l_{\mu\nu}^T+\partial_{\mu}A_{\nu}^T+\partial_{\nu}A_{\mu}^T+\left(\partial_{\mu}\partial_{\nu}-\frac{1}{4}\Box\eta_{\mu\nu}\right)\mu+\frac{1}{4}\lambda\eta_{\mu\nu},
\end{equation}
imposes the Lorentz-like conditions on the new vector and tensor fields: 
\begin{equation}\label{MCOVproperties1}
    \partial_{\mu}l^{T\mu}_{\nu}=0,\qquad \text{and}\qquad \partial_{\mu}A^{T\mu}=0,
\end{equation}
and, in addition, requires:
\begin{equation}\label{MCOVproperties2}
    l^{T\mu}_{\mu}=0,
\end{equation}
one will conclude that the theory propagates one scalar degree of freedom around flat space-time for the linearized theory, as shown in \cite{Alvarez-Gaume:2015rwa}. If, in contrast, one decomposes the metric perturbations into the scalar, vector, and tensor modes:
\begin{equation}\label{decompositionCPT}
    \begin{split}
        &h_{00}=2\phi\\
        &h_{0i}=B_{,i}+S_i,\qquad\qquad S_{i,i}=0\\
        &h_{ij}=2\psi\delta_{ij}+2E_{,ij}+F_{i,j}+F_{j,i}+h_{ij}^{T},\qquad\qquad F_{i,i}=0,\quad h_{ij,i}^{T}=0,\quad h_{ii}^{T}=0,
    \end{split}
\end{equation}
one will find that one of the scalars is constrained, and makes the full action vanish after we solve its constraint and substitute it back into the action \cite{Hell:2023mph}. 

More recently, manifestly covariant approach was claimed to lead to correct number of degrees of freedom if one considers a linearized theory coupled to a source in \cite{Karananas:2024hoh}.  However, the same trick is not applicable if one considers only the free theory, and it does not seem to give agreement beyond the linear order.  As shown in \cite{Hell:2023mph} at the cubic order for the pure $R^2$ gravity in flat space-time, and later confirmed in \cite{Barker:2025gon} and extended to all orders in the perturbation theory, the number of degrees of freedom is still absent -- in contradiction to the claims in \cite{Karananas:2024hoh}.

\subsection{Choosing a background}

The examples so far include fields whose background value vanishes. However, it may happen that one considers the case when there is a non-vanishing background solution, and perturbs around this solution to find the degrees of freedom. This case is often encountered when studying theories of gravity. 

To illustrate this point, let us consider the following action of a scalar field in a gravitational background: 
\begin{equation}\label{scalarField}
    S=\int d^4x\sqrt{-g}\left[\frac{\Mpl^2}{2}R-\frac{1}{2}\nabla_{\mu}\sigma\nabla^{\mu}\sigma-V(\sigma)\right].
\end{equation}
One example of a space-time in which this theory can be studied is the cosmological background. In this case, we can assume that the metric takes the form:
\begin{equation}
    ds^2=-dt^2+a(t)^2\delta_{ij}dx^idx^j,
\end{equation}
where $N(t)$ is the lapse.  The scalar field can also take some non-vanishing background value:
\begin{equation}
    \phi=\phi(t). 
\end{equation}
To find the equations of motion, we can substitute the above quantities into the action (\ref{scalarField}). Then, by varying with respect to the lapse, and subsequently setting it to unity $N=1$, we find:
\begin{equation}
    -V +6\Mpl^{2} H^{2}-\frac{\dot{\sigma}^{2}}{2}=0,
\end{equation}
where $H=\frac{\dot{a}}{a}$ is the Hubble parameter. 
By varying with respect to the scale factor, we find the acceleration equation:
\begin{equation}
    -V+2\Mpl^{2} H^{2}+4\Mpl^{2} \left(H^{2}+\dot{H}\right)+\frac{\dot{\sigma}^{2}}{2 }=0
\end{equation}
Finally, by varying with respect to the scalar field, we find:
\begin{equation}
   V_{,\sigma}+3H \dot{\sigma}+\Ddot{\sigma}=0
\end{equation}

We can see from the above that we have two choices: either we can select that $\sigma(t)=0$, or that $\sigma(t)\neq0$, with the evolution described by the above equations. To count the dof, one can then perturb the metric and the scalar around the background solutions:
\begin{equation}
    g_{\mu\nu}=g_{\mu\nu}^{(0)}+\delta g_{\mu\nu}\qquad\text{and}\qquad \sigma=\sigma^{(0)}+\delta\sigma
\end{equation}
where $\sigma^{(0)}$ and $g_{\mu\nu}^{(0)}$ satisfy the background equations of motion at all times, and expand the action up to second order in these perturbations. However, one has to specify the equations that the background satisfies at any point -- for example, if $\sigma$ vanishes, or not. Choosing the background can lead to significantly different results in the general case. For example, if one considers a vector field instead of the above scalar, with non-minimal coupling to gravity:
\begin{equation}
    S=\int d^4x\sqrt{-g}\left(\frac{\Mpl^2}{2} R-\frac{1}{4}F_{\mu\nu}F^{\mu\nu}-\frac{m^2}{2}A_{\mu}A^{\mu}-\xi RA_{\mu}A^{\mu}\right), 
\end{equation}
and studies this theory in the homogeneous and isotropic Universe, or an anisotropic one, one will find that the choice of the temporal component $A_0(t)=0$ leads to five degrees of freedom -- a scalar, two vector modes, and two tensor modes. If, in contrast, one allows $A_0(t)\neq0$, there will be an additional scalar mode in the theory \cite{DeFelice:2025ykh}.

\section{The Recipe }

So far, we have studied basic field theories and have seen that to find the dof, one has to deal with several different challenges, ranging from the ways to study theories when the system has constraints, gauge redundancy, or a non-trivial background. Importantly, such challenges appear in most of theories, even those involving only a scalar field, and can complicate the overall analysis or even lead to incorrect results. In the following, we will provide a recipe to lead one to the propagating degrees of freedom -- the building blocks of theories -- and overcome these challenges in a straightforward manner. First, we will formulate the most basic case, which assumes that theories include no higher-order derivatives and contain only holonomic constraints, and then expand and discuss also the case when these assumptions are relaxed.

\subsection{The basic recipe}
Suppose that you are given a theory, and you wish to calculate the propagating modes. Then, to find them, we provide the following \textit{recipe}:

\textsc{\textbf{(A)} Specify the background}
\\
One of the main reasons to study any theory is to describe physical phenomena, such as, for example, the acceleration of the Universe, black-holes, or cosmological defects. These are the solutions to the background equations of motion, found by varying the action with respect to fields. 
Notably, depending on the background, the number of degrees of freedom can change. Therefore, when performing the analysis to calculate the dof, it is best to assume that the background is always satisfied. One can ensure this by simply assuming that the background equations of motion always hold at any step in the computation. 

\textsc{\textbf{(B)} Prepare the perturbations}
\\
Once you specify the background (which can also be trivial), it's time to perturb around it. These perturbations will provide the notion of the degrees of freedom. To find them, there are several preparatory things to consider before studying their corresponding action. In particular, it is useful to answer the following questions to decide on the best way forward:
\begin{enumerate}[label=\textit{(\roman*)}]
    \item \textit{Does your system have any symmetries?}
\\ Here, by symmetries, we mean what is left after one separates the time and space components. If then the answer is yes, this might suggest the form of your perturbations, from a general study of components of your field to a much simpler form. For example, in space-times that are homogeneous and isotropic, it is possible to do a decomposition of the modes according to the group of spatial rotations, and decompose the fields into scalars, vectors, and tensors. If the space-time has broken anisotropies in one direction and three spatial dimensions, one can instead decompose the components of the fields into even and odd modes.
 \item \textit{Does your system have any gauge redundancy?}
\\
If yes, analyzing first how the perturbations are transforming under gauge redundancy could be very beneficial: if one fixes the gauge such that it matches the gauge invariant variables, it can decrease the length of the computation. Other gauges are possible too, as long as one makes sure that the gauge is fully fixed.   
\end{enumerate}
It should be stressed that the above two questions can be avoided -- it is perfectly fine if one considers all components of the fields, and does not even fix the gauge. These two decisions would lead to equivalent results, but, also can make the computation significantly longer. 

\textsc{\textbf{(C)} Expand the action }
\\
Once one decides on the background and the type of perturbations around it, it is time to study the perturbations themselves. This can be analyzed either on the level of action or directly on the level of equations of motion. However, studying them directly from the action is more beneficial, since one can then have a clear view into the conditions under which the degrees of freedom are well behaved, such as the positivity of the factors multiplying their kinetic terms, necessary to avoid the ghost modes\footnote{See Appendix for the definition of the ghost dof.}. 

In general, one is supposed to expand the action to the required order in perturbations. If there is a non-vanishing background for the fields, the leading order corresponds to the equations of motion, which are subsequently satisfied. Thus, at first order, the corrections vanish. The kinetic terms for the perturbations are then at second order in the action, and one can also study higher orders to analyze the non-linear corrections. Note that this also holds if the background is trivial, eg, including vanishing background values of the fields, together with the Minkowski space-time. 

\textsc{\textbf{(D)} Analyze the theory }
\\
Once the action is ready, it is time to analyze the modes. Let us first assume that there are at most two time derivatives acting on the fields, and that all constraints are holonomic. Then, essentially, the procedure to find the propagating modes is given by the following: 
\begin{enumerate}[label=(\textit{\roman*})]
    \item Locate the fields that are not propagating -- the constrained modes. 
 \item Find their corresponding constraint by varying the action with respect to them. 
 \item Solve it, and substitute it back into the action. Note that in the case of non-linearities, this is usually done perturbatively. 
 \item Check if the determinant of the kinetic matrix for the Lagrangian density $\mathcal{L}$ of fields $\{\phi_i\}$: 
 \begin{equation}
    \mathcal{L} 
=  K_{ij}(\phi) \dot{\phi}_i\dot{\phi}_j+\text{remaining terms}
 \end{equation}
is vanishing. If yes, this means that there are more non-propagating fields. Locate them in the Lagrangian density. If all of them appear with the form $\dot{\phi}_i\dot{\phi}_i$, this means that one should perform a substitution, which will render one of them to be non-propagating, and then one can repeat the procedure \textit{(i) -- (iv)}. 

\item If, after repeating the procedure, the determinant of the kinetic matrix is not vanishing, and your system contains no higher derivatives, then  you have found the Lagrangian density in terms of the propagating modes. 
\end{enumerate}

\subsection{Higher-order derivatives}
The above procedure works for a wide range of theories. However, one can also easily encounter theories that contain higher-order derivatives. Luckily, this does not drastically complicate the analysis. The key to analyzing it is to reduce the order of derivatives until one reaches maximally two time derivatives per term in the Lagrangian. A simple way to do this is to apply constraints. 

Let us demonstrate this on a simple example for a fourth-order scalar field:
\begin{equation}
    \mathcal{L}=\Box\phi\Box\phi=\ddot{\phi}\ddot{\phi}-2\ddot{\phi}\Delta\phi+\Delta\phi\Delta\phi. 
\end{equation}
The first term is the higher-derivative term, which we now want to reduce to second order. To do this, let's introduce another field:
\begin{equation}
    \mathcal{L}=2\sigma\ddot{\phi}-\sigma^2-2\ddot{\phi}\Delta\phi+\Delta\phi\Delta\phi.
\end{equation}
By varying with respect $\sigma$, we find: 
\begin{equation}
    \sigma=\ddot{\phi}.
\end{equation}
Substituting this back into the action, we return to the starting expression. However, now we have two fields with at most two time derivatives, instead of one with higher-order terms. Therefore, we can easily compute the kinetic matrix, and find that its determinant is different than zero, describing thus two degrees of freedom. In some more complicated systems, it may turn out that the reduction of the order leads to further constraints. In this case, one simply needs to further perform the basic recipe to find the propagating modes. On the other hand, the above trick is applicable also to more complicated systems with an even higher number of time derivatives -- one should, in this case, just introduce the constrained fields, until one removes all second-order derivatives.

Introducing $\sigma$ in this way is just an auxiliary-field rewrite of the standard Ostrogradsky construction \cite{Ostrogradsky:1850fid, Woodard:2015zca}; it is an invertible change of variables in phase space and therefore does not change the number of degrees of freedom, which remains two for a non-degenerate fourth-order scalar. Therefore, the Lagrangians before and after substitution are equivalent. One should note, however, that if the Lagrangian is degenerate, or if substituting $\sigma$ back requires differential or nonlocal relations, or if boundary conditions are not treated consistently, the equivalence fails.

\subsection{Non-holonomic constraints}

In general, our recipe does not apply to the non-holonomic constraint -- constraints which involve both the field and its time-derivatives and cannot be solved only for the fields. However, such constraints, together with the action, can involve the constrained field only through its time derivatives (and possibly spatial derivatives), and not undifferentiated. In that case, one can solve the constraint for the corresponding time derivative of the field and safely, substitute it back into the Lagrangian without changing the number of degrees of freedom.

Let us demonstrate this with two examples. First, we can consider the following Lagrangian:
\begin{equation}\label{L1nhc}
    L_1=\frac{1}{2}\left(\dot{x}^2+\dot{y}^2\right)+\lambda(t)\left(\dot{y}-\ddot{x}\right). 
\end{equation}
In the above expression, $\lambda$ is the Lagrangian multiplier, which gives rise to the following constraint:
\begin{equation}
\dot{y}-\ddot{x}=0. 
\end{equation}
Although the above constraint cannot be algebraically solved for $y$, it can be solved for its derivative. Moreover, this is sufficient, because only the generalized velocity associated with $y$ appears in (\ref{L1nhc}). Therefore, by substituting 
\begin{equation}
    \dot{y}=\ddot{x}
\end{equation}
into (\ref{L1nhc}), we find two degrees of freedom:
\begin{equation}
    L_1=\frac{1}{2}\dot{x}^2+\frac{1}{2}\ddot{x}^2. 
\end{equation}
However, if instead one had:
\begin{equation}
    L_2=\frac{1}{2}\left(\dot{x}^2+\dot{y}^2\right)+\lambda(t)\left(\dot{y}-\ddot{x}\right)-y^2 
\end{equation}
the recipe cannot be applied, because there is a coordinate appearing, which can introduce extra time derivatives $y=y(x,\dot{x})$, and thus it is better to resort to the Hamiltonian approach. 

\subsection{The key message}

The previous recipe can be applied to most of the field theories, theories of gravity, or even physical theories beyond these. However, it also brings an important message when speaking about the degrees of freedom in a theory:
\begin{center}
   \textit{ The number of degrees of freedom depends on the background.   }
\end{center}
It is important to note that not all fields do not necessarily have to be treated on an equal footing for the recipe to be applicable. Let us demonstrate this with an example involving two scalar fields:
\begin{equation}\label{Largder}
     \mathcal{L}=-(1+\sigma)\frac{1}{2}\partial_{\mu}\psi\partial^{\mu}\psi-g\psi\sigma,
\end{equation}
where $g$ is the coupling constant. By varying with respect to $\psi$, we find: 
\begin{equation}\label{psieom}
    \Box\psi+\partial_{\mu}(\sigma\partial^{\mu}\psi)=g\sigma
\end{equation}
while by varying with respect to $\sigma$ we find: 
\begin{equation}\label{psiconstraint}
    \frac{1}{2}\partial_{\mu}\psi\partial^{\mu}\psi+g\psi=0. 
\end{equation}
Let's now perform perturbation theory:
\begin{equation}
    \psi=\psi^{(0)}+\psi^{(1)}+...\qquad \text{and}\qquad \sigma=\sigma^{(0)}+\sigma^{(1)}+...
\end{equation}
where the leading order fields satisfy the linearised field equations. Then, we find:
\begin{equation}
    \Box\psi^{(0)}=g\sigma^{(0)}\qquad \text{and}\qquad \psi^{(0)}=0. 
\end{equation}
The latter constraint implies that $\sigma^{(0)}=0$. At first order in perturbations, we find: 
\begin{equation}
    \Box\psi^{(1)}=g\sigma^{(1)}-\partial_{\mu}(\sigma^{(0)}\partial^{\mu}\psi^{(0)})\qquad \text{and}\qquad g\psi^{(1)}=-\frac{1}{2}\partial_{\mu}\psi^{(0)}\partial^{\mu}\psi^
    {(0)}. 
\end{equation}
However, since the fields at zero order are vanishing, the same result persists also on this order, and the first-order contributions will vanish as well. Therefore, the above theory will not propagate any degrees of freedom. 

Notably, the above two fields are treated on an equal footing -- both were considered to have comparable values. However, we may also regard $\psi$ as an external field, and perform the perturbation theory only in $\sigma$. In this case, from (\ref{psieom}) it follows that 
\begin{equation}
    \Box\psi=0,
\end{equation}
to the leading order in $\sigma$, thus indicating that we have one propagating mode. However, at the same time, we should not forget that there is also a constraint (\ref{psiconstraint}), independent of $\sigma$, which thus constrains $\psi$. From (\ref{psieom}) it then follows that
\begin{equation}
    \partial_{\mu}(\sigma\partial^{\mu}\psi)=g\sigma 
\end{equation}
and this constrains $\sigma$. Thus, we find again no degrees of freedom. 

\subsection{FAQ}

\textbf{\textit{Question 1:} \textit{Why the recipe?}}

The recipe involves expanding the action in perturbations, finding the constraints, solving them, substituting back, and repeating the process until one is left only with the propagating modes in the theory. Naturally, this may seem as a tedious, long, procedure, especially when one tries to tackle also the non-linearities, corresponding to higher orders in expansion on the level of action. Therefore, one might be inclined to wonder why one even does this procedure. 

For a start, one is guaranteed to find the main building blocks of a theory, free of fictitious modes, which could otherwise appear when using the manifestly covariant approach. This is also equivalent to studying a theory on the level of equations of motion, or performing the Hamiltonian analysis. 

It also has a great benefit in accounting for the key properties for the modes -- information about the possible behaviour in the non-linear regime, and clear information on the conditions when these modes are well behaved (See appendix for more information regarding the basics of properties of dof.)

This information is crucial when analysing theories that involve constraints, because with them, one can often encounter the regimes where the perturbative approach breaks down, or which give rise to ghosts -- pathological modes that classically lead to an instability, and cause violation of unitarity when a theory is quantised \cite{Ostrogradsky:1850fid}. 

For example, suppose that one encounters a kinetic term of the following form after resolving all of the constraints: 
\begin{equation}
    \mathcal{L}=-\frac{1}{2}f(t)m^2\partial_{\mu}\chi\partial^{\mu}\chi
\end{equation}
The condition that the field $\chi$ is well behaved is that $f(t)>0$, which is not obvious while working directly with the equations of motion, although the equivalent conditions could also be found by studying the Hamiltonian. Another aspect is that the kinetic term vanishes when $f(t)\to0$, or $m\to0$. This usually indicates strong coupling -- the breakdown of the perturbative approach, at which the non-linear terms containing higher orders of scalar, such as $\mathcal{O}(\chi^3)$ would become comparable to the kinetic term. 

Moreover, having only the propagation fields at hand makes quantisation of a theory especially easy -- one simply finds them, together with the corresponding conjugated momenta, promotes them to operators and imposes canonical commutation relations. From there, it is then easy to find also the path integral, without the need of introducing additional fields, such as the Faddeev-Popov ghosts.

Therefore, having a direct insight into the propagating modes is a powerful tool for gaining information about theories, possible to achieve by following the recipe. 

\noindent\textbf{\textit{Question 2:} \textit{I considered a canonical scalar field and found that its action vanishes when one substitutes the equation of motion for the scalar field. Does this imply that there are no degrees of freedom?}}

No, it does not. It is important to note that our recipe involves only substituting constraints, and not the equations of motion. The latter would clearly lead to wrong results and should not be done at the level of action.

\noindent\textbf{\textit{Question 3:} \textit{How can we know if the gauge is fully fixed?} }

Simply put, if the gauge parameter, which we use to fix the gauge choice, is fully determined, then the gauge choice is fixed. 

Let's illustrate this on two simple examples, getting back to electrodynamics. We have seen that this theory has gauge redundancy: 
\begin{equation}
    A_{\mu}\to\Tilde{A}_{\mu}=A_{\mu}+\partial_{\mu}\lambda.
\end{equation}
Let us then first choose $\lambda$ such that $\Tilde{A}_0=0$. This is known as the Weyl gauge, and it corresponds to setting 
\begin{equation}
    \dot{\lambda}=-A_0
\end{equation}
However, the choice of $\lambda$ that achieves this is not unique, because if for one choice of $\lambda$ we have $A_0=0$, then this also holds for another choice  
\begin{equation}
   \Tilde{\lambda}=\lambda+C_1(\vec{x}). 
\end{equation}
where $C_1$ is the constant of integration. 
As a next example, let us choose the Coulomb gauge, for which $\Tilde{A}_{i,i}=0$. This corresponds to setting
\begin{equation}
    \Delta\lambda=A_{i,i}. 
\end{equation}
Thus, there is only one choice: $\lambda=0$, which corresponds to the vanishing of the divergence of the spatial 3-vector\footnote{Note that we impose the boundary conditions corresponding to the (spatial) Fourier transform of the fields.}. Notably, this also corresponds to the gauge invariant variable -- the transverse modes. 

A similar method can also be applied for cosmological perturbations. See eg \cite{Mukhanov:2005sc} for an excellent overview of gauges in Cosmology.

\noindent\textbf{ \textit{Question 4: What to do if the system is too complicated?}}

Often, it can happen that the system is too complicated to analyze directly. A well-known example is the $f(R)$ gravity. While one can apply the analysis directly in the Jordan frame, in which this theory is formulated, it is much more convenient to go to the Einstein frame, in which the previous theory reduces to Einstein gravity coupled to the scalar field \cite{Whitt:1984pd, Maeda:1987xf, Barrow:1988xh, Teyssandier:1983zz, Sotiriou:2008rp, DeFelice:2010aj}. This is even applicable if the theory is formulated as a function of the square of the Weyl tensor, such as $f(W^2)$ \cite{Hell:2025wha}, or applicable to other theories too. Nevertheless, while these transformations do simplify the analysis, they do not always hold. One should be careful with singular points as these can lead to a different number of degrees of freedom, and should thus be analysed in the original frame (See eg \cite{Hell:2023mph, Hell:2025lbl}) .

\noindent\textbf{ \textit{Question 5: Why do I have to specify the background at the beginning?}}

When studying the perturbations in the case of a non-trivial background, one should always assume that the background is satisfied (only the background equations, and not the equations of motion for the perturbations). This is because the background equations can lead to the cancellation of certain terms, and thus modify the constraints for the perturbations. Therefore, it is not enough to just substitute the background solutions at the end of the computation, but one should impose them at every step on the way. 

\noindent\textbf{ \textit{Question 6: Can this recipe also apply to non-holonomic constraints?}}

In general --  no. In this case, one should resort to other methods, such as the Hamiltonian analysis. However, there are some special cases where the recipe applies. To demonstrate this, let's consider first the following Lagrangian: 
\begin{equation}
    L=\frac{1}{2}\left(\dot{x}^2+\dot{y}^2\right)+\lambda\left(\dot{y}-x\dot{x}\right). 
\end{equation}
If we vary with respect to $\lambda$, which is a Lagrange multiplier, we find:
\begin{equation}
    \dot{y}=x\dot{x}.
\end{equation}
By substituting this back into the Lagrangian, we find the following expression:
\begin{equation}
    L=\frac{1}{2}\left(1+x^2\right)\dot{x}^2, 
\end{equation}
which yields one degree of freedom. Therefore, even though the above constraint was non-holonomic, we managed to use it and apply the recipe because $y$ does not appear anywhere in the action, but only its first derivatives. If one instead also adds another term:
\begin{equation}
    L=\frac{1}{2}\left(\dot{x}^2+\dot{y}^2\right)+\lambda\left(\dot{y}-x\dot{x}\right)-y^2,
\end{equation}
one could encounter a problem, because we do not have an explicit solution for $y$ in general. While for the above constraint, we do, and know that $y$ cannot affect the kinetic matrix, but if we have a more complicated expression, this changes: 
\begin{equation}
    L=\frac{1}{2}\left(\dot{x}^2+\dot{y}^2\right)+\lambda\left(\dot{y}-\ddot{x}\sin{x} \right)-\alpha y^2.
\end{equation}
The constraint is now given by: 
\begin{equation}
    \dot{y}=\ddot{x}\sin{x} . 
\end{equation}
If we set $\alpha=0$, then we can easily obtain the Lagrangian, despite having non-holonomic constraints: 
\begin{equation}
    L=\frac{1}{2}\dot{x}^2+\frac{1}{2}\ddot{x}^2\left(\sin x\right)^2
\end{equation}
which clearly has two degrees of freedom, due to the fourth-order time-derivatives. However, if $\alpha\neq 0$, this is not certain, because we do not know the form of $y=y(x)$. In this case, one should thus rather resort to the Hamiltonian analysis, or analyze the system on the level of equations of motion. 

\noindent\textbf{ \textit{Question 7: Why shouldn't I use the Hamiltonian formalism from the start? }}

We would like to stress here that we do not advocate that one should not use the Hamiltonian formalism. Rather, we are pointing out that using the recipe is much faster -- check Section 6.

\section{A worked out example: The cosmological perturbation theory}
Let us now consider the example of a scalar field with a potential in a cosmological background, and calculate the degrees of freedom following the recipe. We will consider the following action: 
\begin{equation}\label{scalarFieldExample}
    S=\int d^4x\sqrt{-g}\left[\frac{\Mpl^2}{2}R-\frac{1}{2}\nabla_{\mu}\sigma\nabla^{\mu}\sigma-V(\sigma)\right].
\end{equation}

\noindent\textsc{\textbf{(A)}} Let us first specify the background. We will be interested in the non-vanishing solutions of both the background, which we will assume to be homogeneous and isotropic:
\begin{equation}
    ds^2=-dt^2+a(t)^2\delta_{ij}dx^idx^j,
\end{equation}
and for the scalar field $\sigma(t)$. The constraint equations and the equations of motion are then given by:  
\begin{equation}
    -V +6\Mpl^{2} H^{2}-\frac{\dot{\sigma}^{2}}{2}=0,
\end{equation}
\begin{equation}
    -V+2\Mpl^{2} H^{2}+4\Mpl^{2} \left(H^{2}+\dot{H}\right)+\frac{\dot{\sigma}^{2}}{2 }=0,
\end{equation}
and
\begin{equation}
   V_{,\sigma}+3H \dot{\sigma}+\Ddot{\sigma}=0. 
\end{equation}
To study the perturbations, we will solve the above equations in terms of the second derivative of the scale factor, the scalar field, and the potential:
\begin{equation}
   \begin{split}
       V = 6 \Mpl^{2} H^{2}-\frac{\dot{\sigma}^{2}}{2}\qquad\quad
       \Ddot{a}= 
\frac{a \left(\Mpl^{2} H^{2}-\frac{\dot{\sigma}^{2}}{4}\right)}{\Mpl^{2}}\qquad\quad
\Ddot{\sigma}=
-3 H \dot{\sigma}-V_{,\sigma}
   \end{split}
\end{equation}
and assume that they are always satisfied. 

\noindent\textsc{\textbf{(B)}} As the next step, let us prepare the perturbations. We will perturb around the background values for the metric and the scalar field:
\begin{equation}
    g_{\mu\nu}= g_{\mu\nu}^{(0)}+\delta  g_{\mu\nu},\qquad \qquad \sigma=\sigma^{(0)}+\delta  \sigma, 
\end{equation}
where $g_{\mu\nu}^{(0)}$ and $\sigma^{(0)}$ are the background quantities. We can further decompose the metric perturbations according to the group of spatial rotations:
\begin{equation}
    \begin{split}
        \delta g_{00}&=-2\phi\\
        \delta g_{0i}&=a(t)\left(S_i+B_{,i}\right)\\
        \delta g_{ij}&=a^2(t)\left(2\psi \delta_{ij}+2E_{,ij}+F_{i,j}+F_{j,i}+h_{ij}^T\right),
    \end{split}
\end{equation}
where the vector perturbations are divergenceless: 
\begin{equation}
    S_{i,i}=0,\qquad \text{and}\qquad F_{i,i}^T=0. 
\end{equation}
and tensor perturbations divergence-less and traceless: 
\begin{equation}
    h_{ij,j}^T=0\qquad \text{and}\qquad h_{ii}^T=0.
\end{equation}
One can easily see that the linearized theory is invariant under the infinitesimal coordinate transformations by expanding the action up to second order in perturbations:  
\begin{equation}\label{infcoo}
    x^{\mu}\to\Tilde{x}^{\mu}=x^{\mu}+\xi^{\mu}, 
\end{equation}
with 
\begin{equation}
    \xi^{\mu}=(\xi^0, \xi^i),\qquad \xi^{Ti}_{,i}=0,\qquad \xi^i\equiv \delta_{ij} \left( \xi_{j}^T+\zeta_{,j}\right),
\end{equation}

This means that there are four functions which we can use to eliminate four components of the above modes, similarly to the gauge choice in electrodynamics. Under the above transformations, the scalar modes of the metric transform as: 
\begin{equation}
    \begin{split}
        &\phi\to\Tilde{\phi}=\phi-\dot{\xi}^0\qquad \qquad B\to \Tilde{B}=B+\frac{1}{a}\xi^0-a\dot{\zeta} \\
        &E\to \tilde{E}=E-\zeta \qquad\qquad \psi\to \tilde{\psi}=\psi-\frac{\dot{a}}{a}\xi^0  
    \end{split}
\end{equation}
while the scalar field transforms as:
\begin{equation}
    \delta\sigma\to \tilde{\delta\sigma}=\sigma(t)-\dot{\sigma}\xi^0. 
\end{equation}
Using them, we can form the gauge-invariant potentials \cite{Bardeen:1980kt, Mukhanov:1990me}: 
\begin{equation}
  \begin{split}
        \Phi=\phi+\left[a\left(B-a\dot{E}\right)\right]^.,\qquad\Psi=\psi+\dot{a}\left(B-a\dot{E}\right)\qquad \Bar{\sigma}=\delta\sigma-\dot{\sigma}a(B-\dot{E}) .
  \end{split}
\end{equation}
Therefore, by choosing a conformal gauge, defined by:
\begin{equation}
    E=0\qquad\text{and}\qquad B=0,
\end{equation}
The gauge-invariant variables match with those in the conformal gauge, and we can analyse the scalar perturbations in terms of them. 

In addition to the scalar perturbations, we also have the vector and tensor modes. The tensor modes are gauge-invariant by construction, while the vectors transform as: 
\begin{equation}
    S_i\to\tilde{S}_i=S_i-a\dot{\xi}_i\qquad\text{and}\qquad F_i\to\tilde{F}_i=F_i-\xi^T_i.
\end{equation}
Clearly, the following combination is gauge-invariant:
\begin{equation}
    V_i=S_i-a\dot{F}_i. 
\end{equation}
Therefore, we can apply the Poisson gauge:
\begin{equation}
    F_i=0, 
\end{equation}
in which the gauge-invariant variable matches with $S_i$.

\noindent\textsc{\textbf{(C)}} As a next step, we expand the action up to second order in the perturbations. At the leading order, the scalar, vector, and tensor perturbations decouple from each other. Let us focus thus only on the scalar perturbations. For this, it is convenient to go to the Fourier space: 
\begin{equation}
    X=\int \frac{d^3k}{(2\pi)^{3/2}}X_ke^{ikx}
\end{equation}
where $X$ stands for $\phi$, $\psi$ and $\delta\sigma$. Then, we find the following Lagrangian density, after substituting also the background equations of motion, and suppressing the index $k$: 
\begin{equation}
    \begin{split}
        \mathcal{L}_S&=-6a \left(\frac{a^{2} \delta\sigma \left(\phi+3 \psi\right) V_{,\sigma}}{6}+\frac{V_{,\sigma\sigma} \delta\sigma^{2} a^{2}}{12}-3 \dot{a} \left(\dot{\phi}-\frac{4 \dot{\psi}}{3}\right) \left(\phi-\psi\right) \Mpl^{2} a\right.\\&\left.+\left(\frac{\dot{\sigma}^{2} \psi^{2}}{8}+\left(-\frac{3 \dot{\sigma}^{2} \phi}{4}-\Mpl^{2} \ddot{\psi}-\frac{\dot{\sigma} \dot{\delta\sigma}}{2}\right) \psi+\frac{7 \dot{\sigma}^{2} \phi^{2}}{24}+\left(\Mpl^{2} \ddot{\psi}+\frac{\dot{\sigma} \dot{\delta\sigma}}{6}\right) \phi+\Mpl^{2} \dot{\phi} \dot{\psi}-\frac{\dot{\delta\sigma}^{2}}{12}\right) a^{2}\right.\\&\left.-\frac{\left(k^{2}+\frac{9 \dot{a}^{2}}{2}\right) \Mpl^{2} \psi^{2}}{3}-\frac{2 \left(k^{2}-\frac{27 \dot{a}^{2}}{2}\right) \phi \Mpl^{2} \psi}{3}-\frac{7 \phi^{2} \Mpl^{2} \dot{a}^{2}}{2}+\frac{\delta\sigma^{2} k^{2}}{12}\right) 
    \end{split}
\end{equation}

\noindent\textsc{\textbf{(D)}} Let us now perform the analysis, starting with the scalar perturbations.

After performing several integrations by parts, we can notice that the scalar $\phi$ is not propagating. It satisfies the constraint: 
\begin{equation}
    12 \Mpl^{2} a \dot{a}\dot{\psi}-12 \phi \Mpl^{2} \dot{a}^{2}+4 \psi \Mpl^{2} k^{2}+\dot{\sigma}^{2} \phi a^{2}-a^{2} \delta\sigma V_{,\sigma}-a^{2} \dot{\delta\sigma} \dot{\sigma}=0.
\end{equation}
We can solve this equation for $\phi$, and substitute it back into the Lagrangian density, which afterward becomes a function only of $\delta\sigma$ and $\psi$. After performing integrations by parts, we can notice that the two scalars both appear with the time derivatives. However, the determinant of their kinetic matrix is still zero, meaning that one of them is constrained. After performing the decomposition: 
\begin{equation}
    \delta\sigma=\delta\sigma_2+\frac{\dot{\sigma}}{H}\psi, 
\end{equation}
we can notice that $\psi$ loses its kinetic term, and thus becomes constrained. By finding its constraint, solving it, and substituting back into the action, we find after performing additional integration by parts:
\begin{equation}
    \mathcal{L}_S=\frac{a^3}{2}\dot{\delta\sigma}_{2k}\dot{\delta\sigma}_{2(-k)}-f(t)\delta\sigma_{2k}\delta\sigma_{2(-k)}, 
\end{equation}
where:
\begin{equation}
   f(t)= \frac{\left(8 \Mpl^{4} \dot{a}^{2} a^{2} V_{,\sigma\sigma}+8 \Mpl^{4} \dot{a}^{2} k^{2}+12 \Mpl^{2} \dot{\sigma}^{2} \dot{a}^{2} a^{2}+8 \Mpl^{2} V_{,\sigma} \dot{\sigma} \dot{a} a^{3}-\dot{\sigma}^{4} a^{4}\right) a}{16 \Mpl^{4} \dot{a}^{2}}
\end{equation}
Therefore, even though we have started with three scalar modes, we have ended up with only one, whose kinetic term is given by the above expression. 

One can also perform a similar analysis for the remaining modes and find two tensor modes and no vector modes in the theory.

\section{Comparison with the Hamiltonian formalism and the first-order action}

Let's now compare the approaches between the Hamiltonian and Lagrangian formalism, by examining electrodynamics, general relativity, and the pure $R^2$ gravity. 

\subsection{Electrodyamics}

First, following \cite{DiracC}, we will perform the Dirac procedure and find the Hamiltonian of electrodynamics. For this, we start with the Lagrangian density: 
\begin{equation}
    \mathcal{L}=-\frac{1}{4}F_{\mu\nu}F^{\mu\nu}=\frac{1}{2}\left[A_{0,i}A_{0,i}-2A_{0,i}\dot{A}_i+\dot{A}_i\dot{A}_i-A_{i,j}A_{i,j}+A_{i,j}A_{j,i}\right]. 
\end{equation}
The conjugate momenta are given by:
\begin{equation}
    \pi^0=0,\qquad \pi^i=\dot{A}_i-A_{0,i}, 
\end{equation}
and satisfy the following Poisson brackets:
\begin{equation}
    \{A_{\mu}(\vec{x}), \pi^{\nu}(\vec{y})\}=\delta^{\nu}_{\mu}\delta(\vec{x}-\vec{y}). 
\end{equation}
The first momentum is a primary constraint. By computing its Poisson brackets with the Hamiltonian, which is given by:
\begin{equation}
    H=\frac{1}{2}\int d^3y\left(\pi^i\pi^i-2\pi^iA_{0,i}+A_{j,i}A_{j,i}-A_{i,j}A_{j,i}\right), 
\end{equation}
and requiring that this vanishes, we find a secondary constraint:
\begin{equation}
    \pi^{i}_{,i}=0. 
\end{equation}
The two constraints are first-class, since all of the brackets among them vanish, with the total Hamiltonian density given by:
\begin{equation}
\mathcal{H}=\frac{1}{2}\left(\pi^i\pi^i-2\pi^iA_{0,i}+A_{j,i}A_{j,i}-A_{i,j}A_{j,i}\right)+v_1\pi^0+v_2\pi^{i}_{,i}. 
\end{equation}
We can note that $v_2$ can be redefined to remove $A_0$, which also allows us to drop $\pi^0$. This then leaves us with:
\begin{equation}
    \mathcal{H}=\frac{1}{2}\left(\pi^i\pi^i+A_{j,i}A_{j,i}-A_{i,j}A_{j,i}\right)+v\pi^{i}_{,i}
\end{equation}
Clearly, we then have two degrees of freedom by taking into account the constraint. 

In contrast to this approach, our recipe is slightly faster. One right away notices that $A_0$ is not propagating, and thus satisfies: 
\begin{equation}
    \Delta A_0=\Delta \dot{\chi}. 
\end{equation}
where $\chi$ is the longitudinal mode $A_i=A_i^T+\chi_{,i}$. By substituting this back into the Lagrangian density, we find:
\begin{equation}
    \mathcal{L}=\frac{1}{2}\left(\dot{A}_i^T\dot{A}_i^T-A_{i,j}^TA_{i,j}^T\right). 
\end{equation}

It should be pointed out that this procedure is similar to the approach advocated in \cite{Faddeev:1988qp}, which is based on the first-order action:
\begin{equation}
    \mathcal{L}=\pi^i\dot{A}_i-\frac{1}{2}\left(\pi^i\pi^i-2\pi^iA_{0,i}+A_{j,i}A_{j,i}-A_{i,j}A_{j,i}\right)
\end{equation}
There, one can notice that $A_0$ is the Lagrange multiplier, enforcing:
\begin{equation}
    \pi^i_{,i}=0. 
\end{equation}
Then, by decomposing the conjugated momenta and the vector fields into longitudinal and transverse parts, this implies right away:
\begin{equation}
     \mathcal{L}=\pi^{Ti}\dot{A}_i^T-\frac{1}{2}\left(\pi^{Ti}\pi^{Ti}+A^T_{j,i}A^T_{j,i}\right)=\pi^{Ti}\dot{A}_i^T-\mathcal{H}_C
\end{equation}

\subsection{The pure $R^2$ gravity}

As the next comparison, let's consider the pure $R^2$ gravity in the Minkowski background. To find the degrees of freedom, we will first find the first-order action, following the procedure outlined in \cite{Deruelle:2009zk}, which will also set our notation. First, we can note that the action for the pure $R^2$ gravity can be written in the form:
\begin{equation}
    S=\frac{\beta}{2}\int d^4x R^2=\frac{\beta}{2}\int d^4x \left(g^{\mu\rho}g^{\nu\sigma}\rho_{\mu\nu\rho\sigma}\right)^2-\varphi^{\mu\nu\rho\sigma}(R_{\mu\nu\rho\sigma}-\rho_{\mu\nu\rho\sigma}). 
\end{equation}
In particular, by varying the action with respect to $\varphi^{\mu\nu\rho\sigma}$ we find $R_{\mu\nu\rho\sigma}=\rho_{\mu\nu\rho\sigma}$, from which the latter action reduces to the first one. Next, we turn to the ADM variables \cite{Arnowitt:1962hi}:
\begin{equation}
    n_0=-N,\qquad n_i=0,\qquad n^0=\frac{1}{N},\qquad n^i=-\frac{\beta^i}{N}
\end{equation}
\begin{equation}
g_{00}=-N^2+\beta_i\beta^i,\qquad g_{0i}=\beta_i,\qquad g_{ij}=h_{ij},\qquad g^{00}=-\frac{1}{N^2},\qquad g^{0i}=\frac{\beta^i}{N^2},\qquad g^{ij}=h^{ij}, 
\end{equation}
where the indices are raised and lowered by the metric $h_{ij}$. By defining the extrinsic curvature
\begin{equation}
    K_{ij}=\frac{1}{2N}\left(\dot{h}_{ij}-D_i\beta_j-D_j\beta_i\right)
\end{equation}
and using a bit of algebra, and arriving at the Gauss-Codazzi relations (see eg \cite{Gourgoulhon:2007ue}), we find the following action:
\begin{equation}
    S=\frac{\beta}{2}\int d^4x\sqrt{h}N\left[(\rho-2\Omega)^2-\phi^{ijkl}(R^{(4)}_{ijkl}-\rho{ijkl})+4\phi^{ijk}(R^{(4)}_{ijkn}-\rho{ijk})-2\Psi^{ij}(R^{(4)}_{inkn}-\Omega_{ij})\right], 
\end{equation}
where
\begin{equation}
 \begin{split}
        \rho&=h^{ik}h^{jl}\rho_{ijkl},\qquad \Omega=h^{ij}\Omega_{ij}=h^{ij}n^{\mu}n^{\nu}\rho_{i\mu j\nu}, \qquad \phi^{ijkl}=h^{im}h^{jn}h^{ku}h^{lv}\varphi_{mnuv},\qquad \\\phi^{ijk}&=h^{im}h^{jn}h^{ku}n^{\mu}\varphi_{mnu\mu},\qquad \Psi^{ij}=-2h^{ik}h^{jl}n^{\mu}n^{\nu}\varphi_{k\mu l \nu}, 
 \end{split}
\end{equation}
\begin{equation}
    \begin{split}
        R^{(4)}_{ijkl}&=K_{ik}K_{jl}-K_{il}K_{jk}+R^{(3)}_{ijkl}\\
         R^{(4)}_{ijkn}&=D_iK_{jk}-D_jK_{ik}\\
         R^{(4)}_{inkn}&=-\frac{1}{N}\left(\dot{K}_{ij}-\mathcal{L}_{\beta}K_{ij}\right)+K_{ik}K^{k}_j+\frac{1}{N}D_iD_jN, 
    \end{split}
\end{equation}
and 
\begin{equation}
    \mathcal{L}_{\beta}K_{ij}=\beta^kD_k K_{ij}+K_{ik}D_j\beta^k+K_{jk}D_i\beta^k
\end{equation}
is the Lie derivative along $\beta^i$. We can notice that $\phi^{ijkl}$ and $\phi^{ijk}$ are not propagating. By finding their constraints, resolving them, and substituting back into the action, we arrive at the following expression, up to total derivatives: 
\begin{equation}
  \begin{split}
        S=\beta\int d^4x\sqrt{h}N&\left[\frac{1}{2}(K^2-K_{ij}K^
    {ij}-R^{(3)}-2\Omega)^2-\Psi^{ij}(KK_{ij}+K_{il}K^{l}_j+\frac{1}{N}D_iD_jN-\Omega_{ij})\right.\\
    &\left. +\frac{1}{N}K_{ij}(\dot{\Psi}_{ij}-\mathcal{L}_{\beta}\Psi^{ij})\right]
  \end{split}
\end{equation}

The conjugated momenta corresponding to $\dot{\Psi}^{ij}$ and $\dot{h}_{ij}$ are given by the following expressions:
\begin{equation}
  \begin{split}
        p^{ij}=\frac{\delta \mathcal{L}}{\delta \dot{h}_{ij}}&=\beta\sqrt{h}\left[(K^2-K_{kl}K^{kl}-R^{(3)}-2\Omega)(Kh^{ij}-K^{ij})+\frac{1}{2N}(\dot{\Psi}^{ij}-\mathcal{L}_{\beta}\Psi^{ij})\right.\\&\left.
        -\frac{1}{2}\left(\Psi^{ij}K+\Psi^{kl}K_{kl}h^{ij}+\Psi^{il}K^j_l+\Psi^{jl}K^i_l\right)
        \right]
  \end{split}
\end{equation}
and
\begin{equation}
    \Pi_{ij}=\frac{\delta \mathcal{L}}{\delta \dot{\Psi}^{ij}}=\beta\sqrt{h}K_{ij}
\end{equation}

By substituting these expressions back into the action, we find the first-order action for the pure $R^2$ gravity, with the Lagrangian density that is up to the total derivatives given by:
\begin{equation}
 \mathcal{L}=p^{ij}\dot{h}_{ij}+\Pi^{ij}\dot{\Psi}_{ij}-\mathcal{H}, 
\end{equation}
where 
\begin{equation}
    \mathcal{H}=NC+\beta^iC_i,
\end{equation}
is the Hamiltonian density with
\begin{equation}
    \begin{split}
        C&=\frac{1}{\beta\sqrt{h}}\left[\Psi^{ij}(\Pi\Pi_{ij}+\Pi_{il}\Pi^l_j)+2p^{ij}\Pi_{ij}\right]+\beta\sqrt{h}\left(\frac{1}{2}F^2+\Omega_{ij}\Psi^{ij}-D_iD_j\Psi^{ij}\right)\\
        C_i&=-2\sqrt{h}D_j\left(\frac{p^j_i}{\sqrt{h}}\right)+\Pi_{jk}D_i\Psi^{jk}+2\sqrt{h}D_k\left(\frac{\Pi_{ij}\Psi^{jk}}{\sqrt{h}}\right), 
    \end{split}
\end{equation}
and 
\begin{equation}
    F=\frac{1}{\beta^2h}\left[\Pi^2-\Pi_{ij}\Pi^{ij}\right]-R^{(3)}-2\Omega
\end{equation}

Let us now consider this theory in the flat space-time. The only quantities that will have non-vanishing background values are the metric perturbations and the lapse:
\begin{equation}
    N=1+\delta N\qquad\qquad h_{ij}^T=\delta_{ij}+\delta h_{ij}^T
\end{equation}
and thus, for flat-spacetime, at linearised order, we find:
\begin{equation}\label{R2foact}
   \begin{split}
        \mathcal{L}&=\delta p^{ij}\delta \dot{h}_
        {ij}+\delta \Pi^{ij}\delta\dot{\Psi}_{ij}-\frac{2}{\beta}\delta p^{ij}\delta\Pi_{ij}-\frac{\beta}{2}(\delta h_{ij,ij}-\Delta\delta h_{ii})^2-\beta \delta \Omega_{ij}\delta\Psi^{ij}+\frac{\beta}{2}\delta h_{ii,k}\Psi^{jk}_{,j}\\
        &-\beta\delta N\Psi^{ij}_{,ij}+2\delta \beta^i\delta p_{ij,j} 
   \end{split}
\end{equation}
where indices are raised and lowered with $\delta_{ij}$. In  \cite{Barker:2025gon}, the authors have applied the Dirac-Bergmann approach to show that the Hamiltonian of the pure $R^2$ theory does not describe any degrees of freedom, confirming the result of \cite{Hell:2023mph}. In this section, we will apply the method of \cite{Faddeev:1988qp} to study the first-order action, and then compare it with the Lagrangian approach, done in \cite{Hell:2023mph}. 

Starting with the first-order action in flat space-time (\ref{R2foact}), we will decompose all of the fields according to the spatial rotations:
\begin{equation}
    \begin{split}
        \delta p_{ij}&=P^T_{ij}\chi+\frac{1}{3}\delta_{ij}p+p_{i,j}+p_{j,i}+p_{ij}^T\\
        \delta \Pi_{ij}&=P^T_{ij}\xi+\frac{1}{3}\delta_{ij}\pi+\pi_{i,j}+\pi_{j,i}+\pi_{ij}^T\\
        \delta h_{ij}&=P^T_{ij}\sigma+\frac{1}{3}\delta_{ij}h+h_{i,j}+h_{j,i}+h_{ij}^T\\
        \delta \Omega_{ij}&=P^T_{ij}w+\frac{1}{3}\delta_{ij}\Omega+\Omega_{i,j}+\Omega_{j,i}+\Omega_{ij}^T\\
        \delta \Psi_{ij}&=P^T_{ij}\phi+\frac{1}{3}\delta_{ij}\Psi+\Psi_{i,j}+\Psi_{j,i}+\Psi_{ij}^T\\
    \end{split}
\end{equation}
where, 
\begin{equation}
    P_{ij}^T=\partial_i\partial_j-\frac{1}{3}\delta_{ij}\Delta
\end{equation}
all vector and tensor components are divergenceless: $X_{i,i}=0$, and the tensor modes are traceless. In addition, we will decompose the shift as:
\begin{equation}
    \beta^i=\beta^{Ti}+\partial^i\lambda, \qquad \text{with}\qquad \beta^{Ti}_{,i}=0. 
\end{equation}
The scalar, vector and tensor perturbations decouple from each-other at the leading order, and thus we can study them separately. 

For the tensor modes, we can right away notice that one of the fields is constrained:
\begin{equation}
    \delta \Omega_{ij}^T=0. 
\end{equation}
By substituting it back into the action, we find that the tensor component of $\delta\Psi_{ij}$ becomes a Lagrange multiplier, which satisfies: 
\begin{equation}
    \delta\Psi_{ij}^T=0
\end{equation}
Setting it to zero leads to the vanishing of the whole action. Therefore, there are no propagating tensor modes. 

In the case of the vector modes, one can notice that the shift-constraint $\beta^{Ti}$ sets one of the vector modes right away to zero:
\begin{equation}
    p_i=0. 
\end{equation}
Then, by studying the action, we can notice that there is an additional Lagrange multiplier, $\Omega_{i}$. By varying the action with respect to it, we find that 
\begin{equation}
    \Psi_i=0, 
\end{equation}
and thus the whole action for the vector modes vanishes. 

Finally, for the scalar modes, it is convenient to first look at the lapse and shift constraints, which respectively give rise to the following relations:
\begin{equation}
    \Psi=-2\Delta\Phi\qquad \text{and}\qquad p=-2\Delta\chi. 
\end{equation}
In addition, we can notice that $w$ is constrained. By varying the action with respect to it, we find:
\begin{equation}
    \phi=0. 
\end{equation}
Moreover, then, we find that $\xi$ also becomes a Lagrange multiplier, which gives rise to the following:
\begin{equation}
    \chi=0. 
\end{equation}
By substituting everything into the action, we find:
\begin{equation}
    \mathcal{L}_S=\beta\left(\frac{2}{3}\Delta^2\sigma+\frac{2}{3}\Delta h+2\Omega\right)^2
\end{equation}
However, we can notice that neither of these fields has associated time derivatives, and are thus all constrained. This allows us to set them to zero, meaning that the Lagrangian density of the scalar modes vanishes. 

Let us compare the previous procedure with the Lagrangian approach. For this, we perturb the metric around flat space-time:
\begin{equation}
    g_{\mu\nu}=\eta_{\mu\nu}+h_{\mu\nu}, 
\end{equation}
finding the Lagrangian density, which to the leading order in perturbations becomes:
\begin{equation}
    \mathcal{L}=\frac{1}{2}\left(\partial_{\mu}\partial_{\nu}h^{\mu\nu}-\Box h\right)^2
\end{equation}
By decomposing the field according to the transformations under spatial rotations (\ref{decompositionCPT}), and by employing the conformal gauge, in which the scalar potential matches with the gauge-invariant ones, we find: 
\begin{equation}
    \mathcal{L}=4\beta\left(\Delta\phi+3\ddot{\psi}-2\Delta\psi\right)^2
\end{equation}

We can notice right away that $\phi$ is a constrained field, which satisfies:
\begin{equation}
    \Delta\phi=-3\ddot{\psi}+2\Delta\psi. 
\end{equation}
By substituting this constraint back to the Lagrangian density, we find:
\begin{equation}
    \mathcal{L}=0,
\end{equation}
meaning that there are no degrees of freedom at the leading order. One should note that this approach was claimed to change the theory in \cite{Golovnev:2023zen}. However, the agreement between the two approaches clearly shows that this is not the case, and that the theories are equivalent. Therefore, even if one is given the first-order action from the beginning, and can avoid is derivation and right away compare it to the Lagrangian approach, we can find that in this example, the Lagrangian approach is still a faster tool. This, nevertheless, does not mean that one should forget about the Hamiltonian approach -- this approach is especially useful to count the degrees of freedom for an arbitrary background, but it turns out that if one is interested in a particular one, our recipe provides a much faster alternative. 

\textcolor{Black}{Here, we would in addition like to make a comment. The pure $R^2$ theory has three modes for non-vanishing background values of the Ricci scalar. In contrast, in the Minkowski space-time, it propagates no degrees of freedom. For this reason, there were some claims in the literature that the perturbation theory for this theory breaks down \cite{Karananas:2024hoh, Barker:2025noc}. However, this is not necessarily the case. The non-existence of the perturbations such as gravitational modes simply means that the background is rigid, and not an inconsistency in the theory.}

\section{Discussion}

Determining the number and the behavior of the degrees of freedom is one of the most exciting and important questions to answer for any physical theory. It leads one to understand what are the building blocks of their theories, and lays the foundation for their quantization. However, it can also often pose a challenge. This is very transparent in theories for gravity, where the number of the degrees of freedom depends on a background. 

In this work, we have formulated a practical recipe to answer this question, applicable to various physical theories. In essence, this approach, based on the Lagrangian formulation, allows one to rewrite the theory only in terms of its propagating modes. This is extremely powerful for any theory that has constraints, as those hide its underlying structure. As a result, one can be led to surprizes, that can be easily overlooked in more general approaches. 

One such example is the pure $R^2$ theory. As shown in \cite{Alvarez:2018lrg}, for non-vanishing values background values of the Ricci scalar, the theory can be expressed in the Einstein frame, where the action takes the simple form of a Ricci scalar, together with a scalar field and a cosmological constant. However, for vanishing background values of the Ricci scalar, the transformation between the Einstein frame, and the Jordan frame, where the theory is originally formulated becomes singular. As a result, as shown in \cite{Hell:2023mph}, the theory as no propagating degrees of freedom in the Minkowski background. This means that the Minkowski background is rigid. However, absence of modes in a theory is not inconsistent. For example, a massless three form also has an action which is seemingly non-vanishing. However, it also does not propagate any degrees of freedom. 

We would like to stress that the approach presented in this work does not replace the standard Hamiltonian approach to count the degrees of freedom, based on the Dirac-Bergmann algorithm, or the more mathematically rigorous attempts such as in \cite{Heisenberg:2025fxc}, which are focused solely on the equations of motion. Rather, it provides a practical, quick, and complementary alternative. It might also be more useful, depending on the context. For example, if one is interested in the general number of degrees of freedom for a theory of gravity that holds for any background, then the Dirac -Bergmann approach could be the best route. Another possibility then is to generalize the approach in \cite{Heisenberg:2025fxc}. However, if one aims not only to infer the number of propagating modes, but also behaviour each of them, then our approach offers a better route, especially because there might be backgrounds on which the dof change. Notably, as we have demonstrated in Section 6, this route is also quicker, when compared to the Faddeev-Jackiw approach based on the first-order action. Therefore, by following a simple recipe one can infer the nature of theories in a simple, quick, and straight-forward way.

\section{Acknowledgments}

\textit{ A. H. would like to thank Viatcheslav Mukhanov for numerous especially insightful discussions about the degrees of freedom and thoughts on the draft. Part of this work was done during the workshop \textit{"New Perspectives in Cosmology"}, APCTP-2026-F02, in Pohang, APCTP. A. H. would like to thank APCTP, and Jinn-Ouk Gong for hospitality. 
This work was supported in part by JSPS KAKENHI No.~24K00624, and 
by the World Premier International Research Center Initiative (WPI), MEXT, Japan. The work of D.L. is supported by  the German-Israel-Project (DIP) on Holography and the Swampland.
}

\section{Appendix: The nature of modes}

The free particle and scalar field we have analysed in the previous subsection turn out to be the simplest possible theories that one can encounter in nature. However, a small complication, or departure from the above notions, can have big consequences on the behaviour of the modes\footnote{In the following, we will interchangeably use the notion of modes or degrees of freedom, as it is usually done in the literature. }. In this appendix, we will introduce the basic notions related to the modes. 

The \textit{well-behaved dof} refer to the degrees of freedom whose Hamiltonian is bounded from below. Usually, they correspond to stable systems, meaning that if one slightly perturbs them, the system will go back to its original position. An example of such theories is the previously studied free scalar field, given by the Lagrangian density (\ref{LagrDenScalarFree}). 

However, not all dof are well-behaved. As a first example, we may consider a case when a scalar field is unstable, behaving as a tachyon:
\begin{equation}
    \mathcal{L}=-\frac{1}{2}\partial_{\mu}\phi\partial^{\mu}\phi+\frac{1}{2}m^2\phi^2,
\end{equation}
where $m$ is the mass of the scalar. In this case, the equation of motion is given by:
\begin{equation}
    \left(\Box+m^2\right)\phi=0
\end{equation}
In order to see the tachyonic behaviour, it's more convenient to go to the Fourier space:
\begin{equation}
    \phi(t,x)=\int \frac{d^3k}{(2\pi)^{3/2}}\phi_k(t)e^{ikx},
\end{equation}
After which, the above equation becomes:
\begin{equation}
    \Ddot{\phi}_k+\omega_k^2\phi_k=0,
\end{equation}
for each mode $k$, where 
\begin{equation}
    \omega_k^2=k^2-m^2. 
\end{equation}
The above formulation clearly shows how we can think of a scalar field as an infinite number of harmonic oscillators \cite{Mukhanov:2007zz}. In the standard case, the dispersion relation would be given with the positive sign in front of the mass term:
\begin{equation}
    \omega_k^2=k^2+m^2,
\end{equation}
which would correspond to a standard massive stable scalar field. However, in the previous case, the sign in front of the mass term is positive. This implies that the speed of propagation for the scalar field, defined as:
\begin{equation}
    \omega_k^2=c^2k^2
\end{equation}
In flat space-time, it could be larger than unity. Notably, the Hamiltonian in this case is unbounded from below: 
\begin{equation}
    H=\frac{1}{2}\pi^2+\frac{1}{2}\phi_{,i}\phi_{,i} -\frac{1}{2}m^2\phi^2,
\end{equation}
indicating that the system is unstable. 

Another kind of modes that give rise to instabilities are known as \textit{runaway modes}. These modes are characterised by negative gradients: 
\begin{equation}
    \mathcal{L}=\frac{1}{2}\dot{\phi}^2+\frac{1}{2}\phi_{,i}\phi_{,i}-\frac{1}{2}m^2\phi^2,
\end{equation}
and thus their dispersion relation has negative-k modes:
\begin{equation}
    \omega_k=-k^2+m^2,
\end{equation}
and the Hamiltonian is again unbounded from below. One should note that the above theory violates Lorentz invariance, although this is not generally the case \cite{Capanelli:2024pzd, Hell:2024xbv}. 

The final type of modes we will discuss in this subsection are the ghost modes. Let us consider two key examples to find them. The simplest example of such modes is described by the following Lagrangian density: 
\begin{equation}
    \mathcal{L}=-\frac{1}{2}\partial_{\mu}\phi\partial^{\mu}\phi.
\end{equation}
We can notice that, in contrast to the previous cases, this Lagrangian has a negative overall sign. On the one hand, one might think that this is not an issue -- one simply multiplies the Lagrangian density by a factor of $(-1)$, as the equations of motion will not change, when compared to (\ref{healthyscalareom}). However, it will make a difference if the field is coupled to anything else, like another scalar field, or simply considered in the curved space, where gravity becomes important. The Hamiltonian density for the above theory is given by:
\begin{equation}
    \mathcal{H}=-\frac{1}{2}\pi^2-\frac{1}{2}\phi_{,i}\phi_{,i}
\end{equation}
and is clearly unbounded from below.

\bibliographystyle{utphys}
\bibliography{paper}

\end{document}